# Specific Heat of Twisted Bilayer Graphene

**Denis L. Nika[1,2], Alexandr I. Cocemasov[1] and Alexander A. Balandin[2,\*]**

[1]E. Pokatilov Laboratory of Physics and Engineering of Nanomaterials, Department of Physics and Engineering, Moldova State University, Chisinau, MD-2009 Republic of Moldova

[2]Nano-Device Laboratory, Department of Electrical Engineering and Materials Science and Engineering Program, Bourns College of Engineering, University of California – Riverside, Riverside, California, 92521 USA

## Abstract

We have studied the phonon specific heat in single-layer, bilayer and twisted bilayer graphene. The calculations were performed using the Born-von Karman model of lattice dynamics for intralayer atomic interactions and spherically symmetric interatomic potential for interlayer interactions. We found that at temperature $T$<15 K, specific heat varies with temperature as $T^n$, where $n = 1$ for graphene, $n = 1.6$ for bilayer graphene and $n = 1.3$ for the twisted bilayer graphene. The phonon specific heat reveals an intriguing dependence on the twist angle in bilayer graphene, which is particularly pronounced at low temperature. The results suggest a possibility of phonon engineering of thermal properties of layered materials by twisting the atomic planes.

---

\*Corresponding authors: balandin@ee.ucr.edu (AAB).

Specific heat, $C$, is one of the key parameters that characterize the phonon and thermal properties of materials. It is defined as $C = \delta Q / \delta T$, where $\delta Q$ is the change in energy density of a material when temperature changes by $\delta T$ [1]. In the Debye model the phonon specific heat at low temperature varies with temperature $T$ as $T^3$ in three-dimensional (3D) case, $T^2$ in two-dimensional (2D) case and $T$ in one-dimensional case provided that the linear dependence of the phonon frequency $\omega$ on the phonon wavenumber $q$ is assumed [1]. Graphene, as a strictly 2D system [2-3], provides a unique possibility for investigation of 2D electron and phonon gases [2-6]. Due to its high electron mobility and high thermal conductivity, graphene and few-layer graphene (FLG) are considered promising for applications in field-effect transistors [7], interconnects [8-9], thermal interface materials [10-11], and heat spreaders [12-14].

Phonon energy spectra and thermal conductivity in graphene have been investigated both theoretically and experimentally. Experimental studies of optical phonon spectra were performed with Raman spectroscopy [15-16] while the thermal conductivity measurements were carried out using optothermal [4-6] or electrical self-heating methods [17-18]. Theoretical investigations employed different models of the lattice vibrations, molecular dynamics simulations, density functional method and Boltzmann transport equation approach [13-14, 19-23]. Recently the attention has been shifted to twisted few-layer graphene (T-FLG), which demonstrates intriguing electron [24-27] and phonon properties [28-36]. Several independent experimental Raman studies of twisted graphene found series of phonon peaks in different energy regions, which are absent in FLG without the twist [28-30, 32-37]. These experimental results motivate theoretical research to describe phonons in T-FLG.

While the electronic properties of T-FLG were extensively investigated theoretically [26-27] the theoretical studies of phonons in T-FLG are rare [31-32]. In one reported study the Brenner potential for the intralayer interaction and Lennard-Jones potential for the interlayer interactions have been used to investigate phonon modes in a circular twisted bilayer graphene (T-BLG) with a finite radius $R$ = 0.5 nm − 3 nm [31]. It was found that the rotational angle strongly influences ZO phonon modes. We have theoretically investigated the hybrid folded phonons in twisted bilayer graphene using the Born-von Karman model for the intralayer interactions and Lennard-Jones potential for the interlayer interactions [32]. We have shown that the twist angle affects the low-energy phonons the most while the phonon density of states (PDOS) and phonon average group velocity depend only weakly on the rotational angle [32]. While the thirst theoretical studies of the phonon spectra in T-FLG appeared there have been no reported investigations of thermal properties such as specific heat or thermal conductivity. In this Letter we report the results of the theoretical study of specific heat of T-BLG, which reveal its intriguing dependence on the twist angle.

We consider a bilayer graphene structure with a relative rotational angle $\theta$ between the parallel graphene sheets, and define the initial stacking configuration with $\theta = 0^0$ as AB stacking. To obtain T-BLG we rotate one atomic layer of carbon atoms relative to another layer by an angle $\theta$ as shown in Figure 1 (a). The chosen rotation scheme limits our consideration to rotational angles between $\theta = 0^0$ and $\theta = 60^0$, i.e. between AB and AA stacking configuration. The commensurate structures, i.e. structures with translation symmetry, exist for certain rotational angles only determined by the following condition [38]: $\cos\theta(p,n) = (3p^2 + 3pn + n^2/2)/(3p^2 + 3pn + n^2)$, where $p$ and $n$ are co-prime positive integer numbers. If $n$ is not divisible by 3, the number of atoms in T-BLG unit cell is given by [32]:

$$N = 4\left((p+n)^2 + p(2p+n)\right). \quad (1)$$

The unit cells of T-BLG with larger indices ($p,n$) contain a larger number of carbon atoms. For instance, the unit cell of T-BLG with $\theta(1,1) = 21.8°$ contains the smallest possible number of atoms $N = 28$ while a rotation by $\theta(2,1) = 13.2°$ increases this number to $N = 76$. In order to construct the Brillouin zone (BZ) of the T-BLG with the rotation angle $\theta(p,n)$ one should determine the corresponding reciprocal vectors $\vec{g}_1$ and $\vec{g}_2$ of T-BLG [32]. The size of BZ in T-BLG depends on the rotation angle. In Figure 1 (b) we show BZ of the single layer graphene (blue curves) and T-BLG with $\theta(1,1) = 21.8°$ (red curve), determined using a general method described by us in Ref. [32].

In the harmonic approximation, the equation of motion of an atom in T-BLG is given by [32]:

$$m\omega^2(\vec{q})U_k^i(\vec{q}) = \sum_{k=x,y,z}\sum_{s}\sum_{j}\Phi_{kk'}^{ij}(s)\exp\left(\mathbf{i}\vec{r}^{ij}\vec{q}\right)U_{k'}^j(\vec{q}); \; k' = x,y,z; \; i = 1,2,...,N. \quad (2)$$

Here $\omega$ is the phonon frequency, $\vec{q}=(q_x,q_y)$ is the 2D phonon wave vector, $s$ denotes the nearest atomic spheres of the atom $i$; $j$ denotes the atoms of the atomic sphere $s$; $k$ and $k'$ designate the Cartesian coordinate components, $m=19.9442\times10^{-27}$ kg is the mass of a carbon atom, $\Phi_{kk'}^{ij}(s)$ are the interatomic force constants describing the interaction between the atom $j$ and the atom $i$, located at the center of the atomic spheres, and $\vec{r}^{ij}=\vec{r}^{j}-\vec{r}^{i}$, where $\vec{r}^{i}[\vec{r}^{j}]$ is the radius vector of the atom $i[j]$. The phonon energies and thermal conductivity in graphene depends strongly on the interatomic potentials and their parameters [19, 32, 39]. Therefore, a proper choice of the force constant matrices $\Phi$ is crucial for obtaining the correct phonon frequencies and for accurate calculation of phonon specific heat. We describe the strongly anisotropic interatomic interactions in T-BLG using the Born - von Karman (BvK) model of lattice dynamics for intralayer interaction and a spherical potential for interlayer interactions.

The general form of a BvK force constant matrices describing the interaction of an atom with its $s^{th}$-nearest neighbor in a graphene sheet is [32,40-41]:

$$\Phi(s)=-\begin{pmatrix}\alpha_s & 0 & 0\\ 0 & \beta_s & 0\\ 0 & 0 & \gamma_s\end{pmatrix}. \qquad (3)$$

In our model we take into account 4 nearest atomic spheres and 12 independent force constants [42-43], which were fitted to the experimental in-plane phonon dispersion of bulk graphite [44]. In case of the interlayer coupling, the atomic configuration and force constant matrices are directly dependent on the rotational angle $\theta$. Therefore an interatomic potential, which explicitly takes into account the dependence of the interaction strength on the atomic positions, should be used. One possible candidate is a well-known Lennard-Jones potential [45], which has been used to model interlayer interaction in graphite, FLG and T-BLG [21, 31-32, 46]. However, it was pointed out [32, 46] that the Lennard-Jones potential underestimates the frequencies of the shear vibrations in graphite.

Here we describe the interlayer interaction by a spherically symmetric interatomic potential with the following components of corresponding force constant matrices (FCM):

$$\Phi_{kk'}^{ij} = -\delta(r^{ij})\frac{r_k^{ij} r_{k'}^{ij}}{(r^{ij})^2}, \qquad (4)$$

where $\delta$ is the force constant of the interlayer coupling, which depends only on the distance between interacting atoms. Since the interlayer coupling in graphite is very weak and the interaction strength between atoms strongly decreases with $r^{ij}$, we model the dependence of the force constant $\delta$ on $r^{ij}$ as a decaying exponent: $\delta(r^{ij}) = A\exp(-r^{ij}/B)$ with the two fitting parameters $A$=573.76 N/m and $B$=0.05 nm determined from comparison with the experimental phonon frequencies from Γ-A direction in graphite [47]. The cutoff radius of 1 nm was used for $\delta(r^{ij})$ to simplify summation over $j$ in Eq. (2). In Figure 2 (a) we show the results of our calculations of the phonon energy dispersion in bulk graphite along Γ-A direction calculated using FCMs from Eq. (4) (red curves) and Lennard-Jones potential (black curves) with parameters $\varepsilon$=4.6 meV, $\sigma$=0.3276 nm, taken from Ref. [21]. The blue triangles represent the experimental points from Ref. [47]. The curves, calculated with FCMs, are in excellent agreement with the experimental data while the frequencies of the doubly degenerate lowest phonon branches, calculated with the Lennard-Jones potential, are underestimated. The presented data provides experimental validation for our theoretical approach.

In Figure 2 (b-c) we show the phonon energy spectra in AB-stacked bilayer graphene (panel b) and twisted bilayer graphene (panel c) with $\theta = 21.8°$. The experimental phonon energies of graphite, reproduced from Ref. [44], are indicated by the blue triangles. In the single layer graphene (SLG) there are six phonon branches: in-plane transverse acoustic (TA) and optic (TO), in-plane longitudinal acoustic (LA) and optic (LO) and out-of-plane acoustic (ZA) and optic (ZO) [19-20]. Due to the weak van der Waals interactions in bilayer graphene (BLG) these branches split to the twelve nearly-degenerated branches $TA_1/TA_2$, $TO_1/TO_2$, $LA_1/LA_2$, $LO_1/LO_2$, $ZA_1/ZA_2$, $ZO_1/ZO_2$ [6,18,32] with energies close to those in SLG. From Figure 2 (b) we can conclude that our model provides an excellent agreement between the theoretical and experimental phonon frequencies. An evolution of phonon energies in T-BLG with increasing the angle between graphene layers were recently investigated by us in Ref. [32]. The augmentation of the twist angle leads to a decrease of the Brillouin zone size and increase in the number of phonon branches. The hybrid folded phonons appear in T-BLG due to mixing of phonons from different directions of BZ in regular BLG (see Figure 2 (c)). The energies of these phonons depend on the twist angle. The rotation angle - dependent Raman peaks, associated with hybrid folded phonons have been observed experimentally [28-30, 32-37], confirming the theoretical predictions.

For calculation of the phonon specific heat in T-BLG we use the following formula [1, 48]:

$$c_V(T) = \frac{3N_A}{k_B T^2} \int_0^{\omega_{max}} \frac{\exp(\frac{\hbar\omega}{k_B T})}{[\exp(\frac{\hbar\omega}{k_B T})-1]^2} (\hbar\omega)^2 f(\omega) d\omega, \qquad (5)$$

where $\omega$ is the phonon frequency, $\omega_{max}$ is the maximum phonon frequency, $f$ is the 2D normalized phonon DOS, $T$ is the temperature, $N_A$ is the Avogadro constant, $k_B$ is the Boltzmann constant and $\hbar$ is the Planck constant. The normalized phonon DOS is given by:

$$f(\omega) = g(\omega) \bigg/ \int_0^{\omega_{max}} g(\omega) d\omega, \qquad (6)$$

where $g(\omega)$ is the 2D phonon DOS given by the relation:

$$g(\omega) = \int_{q_x} \sum_{s(\omega,q_x)} \sum_{q_y(s,\omega,q_x)} \left|\frac{\partial\omega(q_x,q_y,s)}{\partial q_y}\right|^{-1} dq_x. \qquad (7)$$

Here $s$ numerates phonon branches. In order to calculate $g(\omega)$ we applied a 200×200 2D grid to a 1/4$^{th}$ part of BZ of T-BLG (shown as a green segment in Figure 1 (b)), and then calculated phonon frequencies for every ($q_x$,$q_y$) point in this grid.

The dependence of specific heat on temperature for SLG and AB-BLG is presented in Figure 3 (a). The experimental values of graphite heat capacity reported in Ref. [49] are also shown for comparison by the blue triangles. The low-temperature heat capacity of SLG is higher than that in graphite due to inequality of 2D and 3D phonon DOS. The difference in the heat capacity of SLG, BLG and graphite diminishes with temperature rise. The heat capacities become identical within 0.01% deviation for $T > 2000$ K. At temperatures $T > 2500$ K, all heat capacities approach the classical Dulong-Petit limit $c_V$ = 24.94 J K$^{-1}$ mol$^{-1}$. At small frequencies, ZA phonons demonstrate a quadratic dispersion $\omega \sim q^2$, leading to $c_V(ZA) \sim T$, while TA and LA phonons possess linear dispersions $\omega \sim q$, resulting in $c_V(LA, TA) \sim T^2$. We found that total heat capacity in SLG varies with temperature as $T^n$, where $n = 1$ for $T \leq$ 15 K; $n = 1.1$ for 15 K $< T \leq$ 35 K; $n = 1.3$ for 35 K $< T \leq$ 70 K and $n = 1.5$ for 70 K $< T \leq$ 240 K. The power factor $n$ increases with temperature not only due to greater contribution of LA and TA phonons but also due to the deviation of ZA phonon frequency dispersion from the quadratic law, leading to nearly linear dependence of ZA PDOS on frequency for $\omega >$ 100 cm$^{-1}$. Our results for heat capacity in SLG are qualitatively similar to those described in Refs. [50-51] with the exception of the power factor values $n = 1$ [50] and $n = 1.1$ [51] for $T <$ 100 K. This discrepancy is explained by the fact that in Refs. [50-51] the authors did not take into account the non-parabolicity of ZA dispersion for $\omega >$ 100 cm$^{-1}$. In AA-BLG or AB-BLG $c_V \sim T^{1.6}$ for wide temperature range $T \leq$ 170 K due to the changes in PDOS in comparison with those in SLG.

In Figure 3 (b) we plot a difference between the specific heat in AB-BLG and T-BLG as a function of temperature: $\Delta c_v(\theta) = c_v(AB) - c_v(\theta)$ for $\theta = 21.8°$, $\theta = 13.2°$ and $\theta = 9.4°$. The change in the specific heat due to twisting is relatively weak in a wide temperature range 20 K - 2000 K. It attains its maximum value ~ 0.028 J K$^{-1}$ mol$^{-1}$ at $T$ ~ 250 K. At the same time, at low temperatures the relative difference between specific heat in AB-BLG and T-BLG $\eta = (1 - c_V(\theta)/c_V(AB)) \times 100\%$ constitutes substantial 10-15% at $T$ = 1 K and ~ 3-6% at $T$ = 5 K in dependence on $\theta$ (see blue, red and green curves from the inset to Figure 3 (b)). The low temperatures specific heat depends stronger on the twist angle because twisting affects the low-frequency ZA phonon modes the most [32]. A somewhat similar effect of modulation of the specific heat was recently reported for SLG under strain [52-53]. The authors found that a stronger modification of the specific heat occurs at temperatures below 1 K. The temperature dependence of low-temperature specific heat in T-BLG with $\theta = 21.8°$ differs from SLG or BLG: $c_V \sim T^{1.3}$ for T < 10 K and $\sim T^{1.6}$ for 10 K $\leq T \leq$ 100 K. One should expect that twisting can produce stronger effects on the specific heat of the T-FLG with the larger number of the atomic planes rotated with respect to each other as well as in van der Waals materials with stronger interlayer coupling.

In summary, we found that the phonon specific heat reveals an intriguing dependence on the twist angle in T-BLG, which is particularly pronounced at low temperature. The results suggest a possibility of phonon engineering of thermal properties of layered materials by twisting the atomic planes. Specifically, the twisting decreased the specific heat in T-BLG by 10 − 15% in comparison with reference BLG at $T \sim 1$ K. It changes the temperature dependence of the heat capacity from $c_V(\text{BLG}) \sim T^{1.6}$ to $c_V(T\text{BLG}) \sim T^{1.3}$ for $T$<10 K. The maximum deviation of the specific heat in TBG from that in AB-bilayer graphene was found at $T \sim 250$ K.

**Acknowledgements**

This work was supported by the National Science Foundation (NSF) grant ECCS-1307671 Two Dimensional Performance with Three Dimensional Capacity: Engineering the Thermal Properties of Graphene. DLN and AIC acknowledges the financial support through the Moldova State projects 11.817.05.10F, 14.819.16.02F and Moldova-STCU project 14.820.18.02.012 STCU.A/5937.

# Figure Captions

**Figure 1:** (a) Rotational scheme. Note that rotational axis *R* passes through atoms, which lie exactly above each other. (b) Brillouin zones of SLG (blue and red hexagons) and T-BLG with $\theta = 21.8°$ (green hexagon). Γ and K denote two high-symmetry points of T-BLG BZ. The basis vectors of the reciprocal lattice of SLG and T-BLG are denoted as $\vec{b}_1, \vec{b}_2$ and $\vec{g}_1, \vec{g}_2$ correspondingly.

**Figure 2:** Phonon energy dispersion in (a) graphite along the $\Gamma - A$ direction, (b) in AB-bilayer graphene, and (c) twisted bilayer graphene with $\theta = 21.8°$. The blue triangles represent experimental data from Ref. [47] (panel a) and from Ref. [44] (panel b). The red lines in Figure 1 (a) correspond to the theoretical results calculated with the force constant matrices from Eq. (3) while the black lines show results calculated using the Lennard-Jones potential.

**Figure 3** (a) Dependence of the specific heat at constant volume on temperature in SLG (gray curve) and AB-BLG (black curve). The blue triangles represent the experimental results for graphite, reported in Ref. [49]. The inset shows the results plotted for a wide temperature range 100 K – 3000 K. (b) Dependence of the deviation $\Delta c_v(\theta)$ of the specific heat in T-BLG from that in AB-BLG on the temperature. The inset shows the relative deviation $\eta$ between AB-BLG and T-BLG specific heats as a function of temperature.

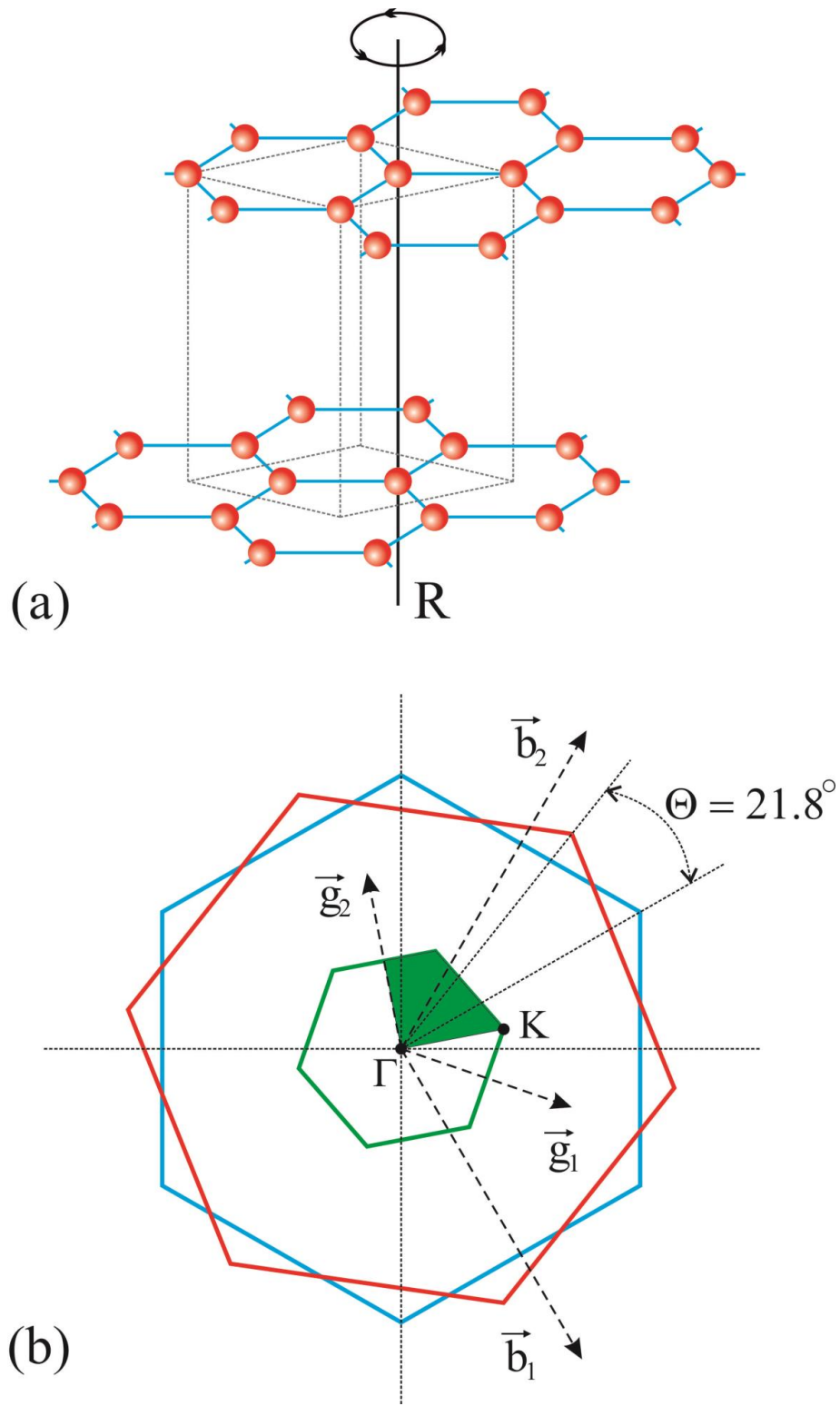

Figure 1

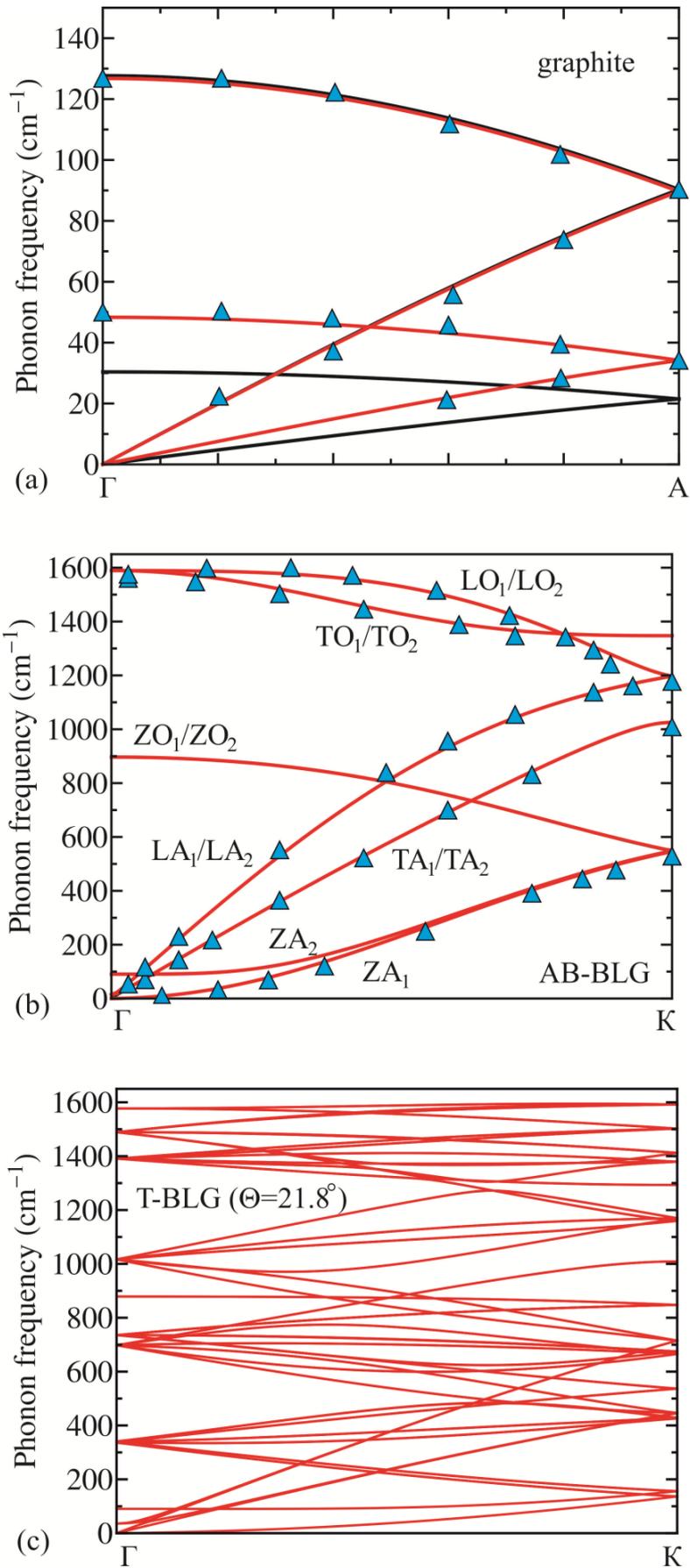

Figure 2

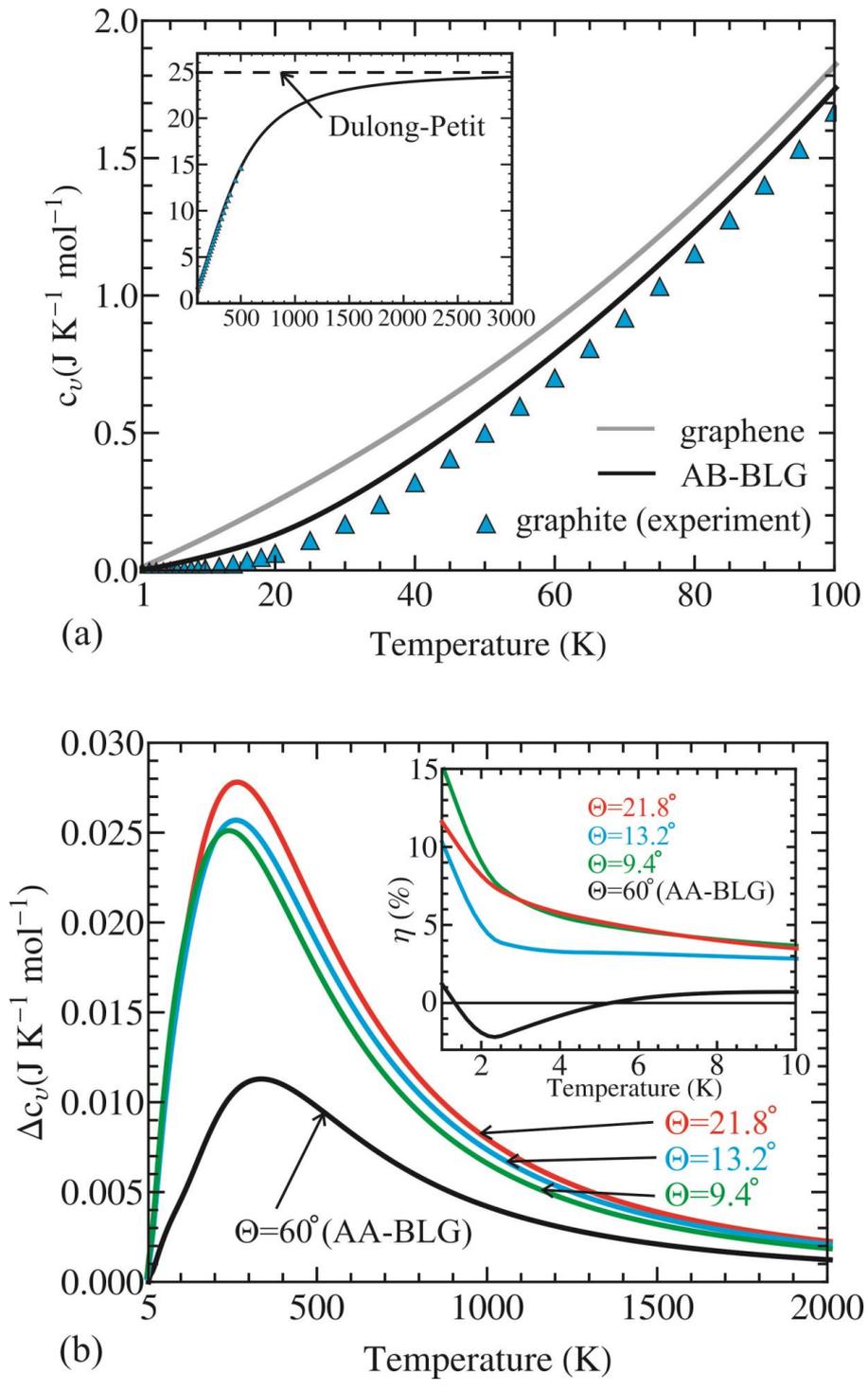

Figure 3